\documentstyle[epsf]{mn}

\begin{document}
\def\lsun{{\rm L_{\odot}}}
\def\msun{{\rm M_{\odot}}}
\def\rsun{{\rm R_{\odot}}}
\title{Black--Hole Transients and the Eddington Limit}
\author[A.R. King]{A.R. King$^1$\\ 
$^1$ Astronomy Group, University of Leicester, Leicester LE1
7RH, U.K}
\maketitle            
\begin{abstract}
I show that the Eddington limit implies a critical orbital period 
$P_{\rm crit}({\rm BH}) \simeq 2$~d beyond which black--hole 
LMXBs cannot appear as persistent systems. The unusual behaviour of
GRO J1655-40 may result from its location close to this critical period.
\end{abstract}

\section{Introduction}
It is now well understood that the accretion discs in low--mass X--ray
binaries (LMXBs) are strongly irradiated by the central X--rays, and
that this has a decisive effect on their thermal stability (van
Paradijs, 1996; King, Kolb \& Burderi, 1996). Irradiation stabilizes
LMXB discs compared with the otherwise similar ones in cataclysmic
variables (CVs) by removing their hydrogen ionization zones.  In CVs
this instability causes dwarf nova outbursts, and in LMXBs it produces
transient outbursts rather than persistent accretion. The irradiation
effect appears to be weaker if the accretor is a black hole rather
than a neutron star, possibly because of the lack of a hard surface
(King, Kolb \& Szuszkiewicz, 1997). The result is that neutron--star
LMXBs with short ($\sim$ hours) orbital periods tend to be persistent,
while similar black--hole binaries are largely transient. Both types
of LMXBs must be transient at sufficiently long orbital periods, since
a long period implies a large disc, so that a large X--ray luminosity
would be needed to keep the disc edge ionized and thus suppress
outbursts. We can write this stability requirement as
\begin{equation}
\dot M_{\rm crit}^{\rm irr} \sim R_{\rm d}^2 \sim P^{4/3},
\label{crit}
\end{equation}
where $\dot M_{\rm crit}^{\rm irr}$ is the minimum central accretion
rate required to keep the disc stable, $R_{\rm d}$ is the outer
disc radius, and $P$ is the orbital period, and we have used Kepler's
law. Thus for large $P$, $\dot M_{\rm
crit}^{\rm irr}$ will rise above any likely steady accretion
rate, making long--period systems transient.
This simple prediction (King, Frank, Kolb \& Ritter, 1997)
seems to be borne out by the available evidence.

\section{The Critical Accretion Rate}
The precise coefficient in (\ref{crit}) depends on uncertainties
in the vertical disc structure (see the discussion in Dubus, Lasota,
Hameury \& Charles, 1999). Here I adopt the form derived by King, Kolb
\& Szuszkiewicz (1997). They argued that for a steady black--hole accretor,
the central irradiating source is likely to be the inner disc rather than a
solid spherical surface, as for a steady neutron--star accretor. (Note
that during an outburst of a {\it transient} black--hole system such a
spherical source may be present, as the central accretor may develop a
corona.) For a small source  
at the centre of the disc and lying in its plane, the irradiation temperature 
$T_{\rm irr}(R)$ is given by
\begin{equation}
T_{\rm irr}(R)^4 = {\eta \dot Mc^2(1-\beta)\over 4\pi \sigma R^2}
\biggl({H\over R}\biggr)^2
\biggl({{\rm d}\ln H\over {\rm d}\ln R} - 1\biggr) 
\label{eq3}
\end{equation}
(Fukue, 1992). Here $\eta$ is the efficiency of rest--mass energy
conversion into X--ray heating, $\beta$ is the X--ray albedo, and
$H(R)$ is the local disc scale height.  The minimum accretion rate
required to keep the disc in the high state is given by setting
$T_{\rm irr}(R) = T_H$, where is $T_H$ is the hydrogen ionization
temperature. Since $T$ always decreases with $R$, the global minimum
value $\dot M_{\rm crit}^{\rm irr}$ is given by conditions at the
outer edge $R_{\rm d}$ of the disc. For the parametrization adopted by King,
Kolb \& Szuszkiewicz (1997), and $\eta = 0.2$, this leads to
\begin{equation}
\dot M_{\rm crit}^{\rm irr}(R) = 2.86\times
10^{-11}m_1^{5/6}m_2^{-1/2}f_{0.7}^2gP_h^{4/3}\ \msun\ {\rm yr}^{-1},
\label{crit2}
\end{equation}
where 
$f_{0.7}$ is the disc filling fraction $f$
(the ratio of $R_{\rm d}$ to the accretor's
Roche lobe) in units of 0.7; $m_1, m_2$ are the accretor and
companion star mass in $\msun$; and 
\begin{eqnarray}
\lefteqn{g = } \nonumber \\ 
\lefteqn{\biggl({1-\beta\over 0.1}\biggr)^{-1}\biggl({H\over
0.2R}\biggr)^{-2}
\biggl({2/7\over {\rm d}\ln H/{\rm d}\ln R - 1}\biggr)
\biggl({T_H\over 6500\ {\rm K}}\biggr)^4}
\label{g}.
\end{eqnarray}
Equation (\ref{crit2}) is the same as eqn (12) of King, Kolb \&
Szuszkiewicz (1997) apart from the factors $f_{0.7}^2g$, there taken
as unity. All of the uncertainties over disc thickness, warping,
albedo etc are lumped into the quantity $g$. With $g \simeq f_{0.7}
\simeq 1$, equation (\ref{crit2}) appears to be largely successful in
predicting that systems with reasonably massive ($5 - 7\msun$) black
holes and main--sequence companions should be transient. By contrast,
neutron star systems with main--sequence companions should be
persistent, as the index of the ratio $H/R$ in (\ref{eq3}) is unity,
implying more efficient disc irradiation. (Equation (\ref{crit2}) also
implies that lower--mass black hole systems might be persistent.) These
results suggest that the quantity $g$ appearing in (\ref{crit2})
cannot be too far from unity.

\section{The Eddington Limit}
Here I concentrate an another aspect of eq. (\ref{crit2}) which
does not seem to have received much attention. Namely, for large
enough $P$, $\dot M_{\rm crit}^{\rm irr}$ must exceed the Eddington
accretion rate
\begin{equation}
\dot M_{\rm Edd} \simeq 1\times 10^{-8}m_1\msun\ {\rm yr}^{-1}.
\label{edd}
\end{equation}
The obvious consequence of eqs. (\ref{crit2}, \ref{edd}) is that for
sufficiently long orbital periods irradiation will be unable to
suppress outbursts, as the required central luminosity exceeds the
Eddington limit, and the system presumably cannot be both
super--Eddington and persistent. Note that this conclusion holds
whatever the {\it actual} value of the mass transfer rate in the
particular binary happens to be.  Thus we should expect to find no
persistent LMXBs above a certain critical orbital period $P_{\rm
crit}$.  For the neutron--star case this was recognised by Li \& Wang
(1998), who found $P_{\rm crit}({\rm NS}) \simeq 20$~d, in agreement
with observation. For the black--hole case, combining (\ref{crit2},
\ref{edd}) gives
\begin{equation}
P_{\rm crit}({\rm BH}) \simeq
2.0f_{0.7}^{-1.5}g^{-0.75}\biggl({\dot M\over 0.5\dot M_{\rm
Edd}}\biggr)^{0.75}m_1^{1/8}m_2^{1/8}~{\rm d},
\label{pcrit}
\end{equation}
where we have included a factor $(\dot M/0.5\dot M_{\rm
Edd})$ to allow for the fact that the radiation pressure limit for the
accretion rate $\dot M$ may in practice be below $\dot M_{\rm Edd}$.
We thus expect to find no persistent black--hole LMXBs above this
period. This is indeed supported by the available data, but hardly
surprising in view of the difficulty in identifying black holes in
persistent systems. Note that in {\it high--mass} black--hole systems
such as Cygnus X--1, the powerful UV luminosity of the companion star,
as well as the small disc size expected in a wind--fed system, are
both likely to keep the disc hot and therefore give a persistent
system.

\section{GRO~J1655-40}
With $g \sim 1$ as argued above, the value of $P_{\rm crit}({\rm BH})$
found above is close to the observed period $P = 2.62$~d of the
black--hole soft X--ray transient GRO J1655-40 (the nearest periods
among alternative black--hole systems are $P = 6.47$~d for V404 Cyg
and $P = 1.23$~d for 4U 1543--47). Indeed Kolb et al (1997) pointed
out the system's proximity to the Eddington limit during outburst, and
Hynes et al. (1998) explicitly suggested that no globally steady disc
solution might be possible for this system with $\dot M < \dot M_{\rm
Edd}$.  GRO~J1655-40 is unusual in at least two respects:

1. The companion star has spectral type F3 -- F6IV and mass $M_2
\simeq 2.3\msun$. On a conventional view, this places it in the
Hertzsprung gap. The companion star should therefore be expanding on a
thermal timescale and thus driving a mass transfer rate $-\dot M_2
\sim 10^{-7}\msun~{\rm yr}^{-1}$ (Kolb et al., 1997).  This is well
above the appropriate value of $\dot M_{\rm crit}^{\rm irr}$, making
it puzzling that the system is nevertheless transient, and far above
the mean mass accretion rate of $\dot M_{\rm obs} = 1.26\times
10^{-10}\msun~{\rm yr}^{-1}$ deduced by van Paradijs (1996) from
observation.  Reg\"os, Tout \& Wickramasinghe (1998) appeal to
convective overshooting to increase the main--sequence radius of stars
of $\sim 2\msun$. The companion might then be on the main sequence
rather than in the Hertzsprung gap. This implies a slower evolutionary
radius expansion, bringing the predicted mass transfer rate below
$\dot M_{\rm crit}^{\rm irr}$.  However $-\dot M_2$ is still predicted
to lie uncomfortably far above $\dot M_{\rm obs}$.

2. The system was first detected in an outburst in 1994, and had
probably been quiescent for at least 30~yr before that. Yet two more
outbursts followed in the next two years.

The considerations given here offer explanations for both of these
unusual features. First, if $P > P_{\rm crit}({\rm BH})$, the system
must be transient in some sense, regardless of the actual mass
transfer rate (cf Hynes et al., 1998). It would therefore not be
necessary to appeal to convective overshooting. Further, since the
system is close to $P_{\rm crit}({\rm BH})$, it is evidently accreting
at a value close to the Eddington rate during its quasi--steady states
(see below), making it natural that $\dot M_{\rm obs}$ is much smaller
than the predicted mass transfer rate $-\dot M_2$.

Second, assuming that the quantity $g$ has a relatively constant value
close to unity, as argued above, we see from (\ref{pcrit}) that the
value of $P_{\rm crit}({\rm BH})$ is most sensitive to the filling
factor $f$ (I consider the effect of dropping the assumption
$g\sim$~constant below). Thus if $f$ decreases, $P_{\rm crit}({\rm BH})$ can
increase above the actual orbital period, allowing irradiation to keep
the disc in the high state (prolong an outburst) for as long as $f$
remains sufficiently small. Hence the unusual outburst behaviour of
GRO~J1655-40 may be explicable in terms of the time evolution of the
disc size. Encouragingly there is some observational evidence (see the
discussion in Orosz \& Bailyn, 1997) that the grazing eclipses seen in
the optical are time--dependent, just as expected if the disc size
varies. Moreover Soria, Wu \& Hunstead (1999) find evidence from the
rotational velocities of double--peaked emission lines that the disc
is at some epochs slightly larger than its tidal limit. The large
resultant torques on the disc suggest that this state cannot persist
and the disc must eventually shrink.

In fact we do expect $f$ to evolve systematically: in the early part
of an outburst, the central accretion of low angular--momentum
material will raise the average disc angular momentum and thus cause
$f$ to increase, hence lowering $P_{\rm crit}({\rm BH})$ and making
the system more vulnerable to a return to quiescence. However at some
stage matter transferred from the companion will tend to reduce the
angular momentum of the outer disc, thus decreasing $f$, raising
$P_{\rm crit}({\rm BH})$ and allowing irradiation to stabilize the
disc in the high state. But eventually the disc must grow towards its
tidal limit, increasing $f$ and thus lowering $P_{\rm crit}({\rm BH})$
again, finally enforcing a return to quiescence.  Obviously a full
disc code is required to follow this sequence in detail and to check
if it can account qualitatively for the unusual outburst behaviour of
GRO~J1655--40. 

Clearly, systematic evolution of one or more of the quantities
appearing in $g$ during the outburst could have a similar effect in
making $P_{\rm crit}({\rm BH})$ oscillate around the actual orbital
period $P$. The most likely alternative candidate is the disc aspect
ratio $H/R$, which would appear explicitly with the power 1.5 if we
substitute for $g$ in (\ref{pcrit}). The aspect ratio could evolve
systematically on a viscous timescale because the disc may warp under
radiative torques (Pringle, 1996). A warp presenting more of the disc
surface to the central source would tend to stabilize it against a
return to quiescence even though the central luminosity was below the
Eddington limit. Again considerably more work is required to check
this possibility.

\section{Conclusions}
I have shown that the Eddington limit implies a critical orbital
period $P_{\rm crit}({\rm BH})$ beyond which black--hole LMXBs cannot
appear as persistent systems. The precise value of $P_{\rm crit}({\rm
BH})$ is subject to uncertainties expressed by the quantity $g$ in
(\ref{crit2}). I have argued that $g$ cannot be very far from unity if
we are to understand the difference in the stability properties of
discs in neutron--star and black--hole systems. In this case
GRO~J1655-40 lies much closer to $P_{\rm crit}({\rm BH})$ than any
other black--hole system.

The unusual behaviour of GRO~J1655-40 may result from its location
very close to $P_{\rm crit}({\rm BH})$; evolution of the disc size or
possible radiative warping may move the system across the boundary
where a sub--Eddington luminosity can keep the disc stably in the high
state. This system, and those at longer orbital periods, probably have
central accretion rates which are highly super--Eddington during
outbursts. Since observed radiative luminosities are mildly
sub--Eddington, most of this mass must be expelled. Strong support for
this comes from the observation of P Cygni profiles in GRO~J1655--40
(Hynes et al., 1998).  The superluminal jets observed (Hjellming \&
Rupen, 1995) in an outburst of this system may therefore simply
represent the most dramatic part of this outflow.

\section{Acknowledgment}
I gratefully acknowledge the support of a PPARC Senior Fellowship.

\end{document}